\documentclass[iop,deluxetable,twocolappendix]{emulateapj}
\usepackage{amsmath}

\newcommand{\GRay}{\texttt{GRay}}

\newcommand{\llangle}{\langle\!\langle}
\newcommand{\rrangle}{\rangle\!\rangle}

\newcommand{\rS}{r_{\rm S}}
\newcommand{\Ne}{n_\mathrm{e}}
\newcommand{\Bt}{\beta_\mathrm{threshold}}
\newcommand{\td}{\theta_\mathrm{disk}}
\newcommand{\tf}{\theta_\mathrm{funnel}}
\newcommand{\Tf}{T_\mathrm{e,funnel}}
\newcommand{\Te}{T_\mathrm{e}}
\newcommand{\Ti}{T_\mathrm{i}}
\newcommand{\DPA}{\Delta\mathrm{PA}}
\newcommand{\sgra}{Sgr~A$^*$}

\begin{document}

\title{The Power of Imaging: Constraining the Plasma Properties of
  GRMHD Simulations using EHT Observations of \sgra}

\author{Chi-Kwan Chan, Dimitrios Psaltis, Feryal \"Ozel}
\affil{Steward Observatory and Department of Astronomy,
  University of Arizona,
  933 N. Cherry Ave., Tucson, AZ 85721}

\author{Ramesh Narayan}
\affil{Institute for Theory and Computation,
  Harvard-Smithsonian Center for Astrophysics,
  60 Garden Street, Cambridge, MA 02138}

\and

\author{Aleksander S{\c a}dowski}
\affil{MIT Kavli Institute for Astrophysics and Space Research,
  77 Massachusetts Ave, Cambridge, MA 02139}

\begin{abstract}
  Recent advances in general relativistic magnetohydrodynamic
  simulations have expanded and improved our understanding of the
  dynamics of black-hole accretion disks.
  However, current simulations do not capture the thermodynamics of
  electrons in the low density accreting plasma.
  This poses a significant challenge in predicting accretion flow
  images and spectra from first principles.
  Because of this, simplified emission models have often been used,
  with widely different configurations (e.g., disk- versus
  jet-dominated emission), and were able to account for the observed
  spectral properties of accreting black-holes.
  Exploring the large parameter space introduced by such models,
  however, requires significant computational power that exceeds
  conventional computational facilities.
  In this paper, we use \GRay, a fast GPU-based ray-tracing algorithm,
  on the GPU cluster \texttt{El~Gato}, to compute images and spectra
  for a set of six general relativistic magnetohydrodynamic
  simulations with different magnetic field configurations and
  black-hole spins.
  We also employ two different parametric models for the plasma
  thermodynamics in each of the simulations.
  We show that, if only the spectral properties of \sgra\ are used,
  all twelve models tested here can fit the spectra equally well.
  However, when combined with the measurement of the image size of the
  emission using the Event Horizon Telescope, current observations
  rule out all models with strong funnel emission, because the funnels
  are typically very extended.
  Our study shows that images of accretion flows with horizon-scale
  resolution offer a powerful tool in understanding accretion flows
  around black-holes and their thermodynamic properties.
\end{abstract}


\section{Introduction}

General relativistic magnetohydrodynamic (GRMHD) simulations of
accretion flows onto black-holes have significantly expanded and
improved our understanding of accretion physics (see
\citealt{2013LRR....16....1A} and \citealt{2014ARA&A..52..529Y} for
recent reviews).
Multiple numerical algorithms have been developed, which allow for
accurate simulations of the turbulent magnetohydrodynamic (MHD) flow
in the curved spacetime near a black-hole, the plunging of matter at
the location of the innermost stable circular orbit, the accretion
through the event horizon, and the generation of outflows and jets.

An important limitation of all current GRMHD simulations is their
inability to follow the thermodynamic properties of the electrons in
the flow.
This is a crucial ingredient in predicting the observational
characteristics of accretion flows onto black-holes, because the
emission from such systems is dominated by the radiative properties of
the electrons.
There are several reasons that hamper rapid progress in this aspect of
global simulations.
First, the heating and, in general, the acceleration of electrons by
the dissipation of the MHD turbulence occurs at sub-grid scales
\citep[see, e.g.,][]{2012ApJ...755...50R}.
Second, the transport of heat by conduction is highly anisotropic and
takes place in a regime that is poorly understood \citep[see,
  e.g.,][]{2008MNRAS.389.1815S}.
Finally, radiative energy losses are characterized by cross sections
that have very strong dependence on photon energy and render
approximate radiative transfer schemes inadequate \citep[see,
  e.g.,][]{2012ApJS..199....9D}.

Improving our understanding of the relevant physical process will be
stimulated in the near future by technological advances in spatially
resolved imaging observations of black-holes.
The Event Horizon Telescope (EHT) will perform mm VLBI observations of
a number of supermassive black-holes and will achieve horizon-scale
resolution for at least two well studied systems, \sgra\ and M87.
Early EHT observations have confirmed that the size of the emitting
region at 1.3\,mm in both systems is at most equal to a few
Schwarzschild radii \citep{2008Natur.455...78D, 2012Sci...338..355D}.
This simple piece of evidence, in combination with the broadband
spectra of the sources, can already place stringent constraints on the
physical conditions and geometries of the accretion flows \citep[see,
  e.g.,][]{2009ApJ...697...45B, 2011ApJ...735..110B,
  2009ApJ...703L.142D, 2010ApJ...717.1092D, 2011ApJ...730...36D,
  2012JPhCS.372a2023D, 2012ApJ...755..133S, Psaltis2014}.
In the near future, incorporating ALMA and the South Pole Telescope to
the array of telescopes that comprise the EHT will generate
polarization dependent images of \sgra\ and M87 at two wavelengths
(1.3\,mm and 0.8\,mm) and at different epochs \citep[see,
  e.g.,][]{2009ApJ...695...59D, 2011ApJ...735...57B}.
This wealth of data will provide the observational foundation against
which the results of GRMHD simulations will be calibrated.

The power of comparing theoretical models to spatially resolved
observations of accreting black-holes becomes apparent when exploring
the relative importance of the shearing flow (or, for simplicity, the
disk) and of the bulk outflow or jet in a system.
Albeit typically absent from simple analytic models of accretion
disks, outflows, such as highly relativistic jets and winds are common
features of GRMHD simulations that often form spontaneously
\citep[see, e.g.,][]{2004ApJ...611..977M}.
The plasma density in an outflow is typically much lower than in the
disk but the magnetic field and relativistic Lorentz factor are much
larger, potentially dominating the emerging radiation from the system.
Indeed, semi-analytic models dominated by a radiatively inefficient
accretion flow \citep[see, e.g.,][]{1998ApJ...492..554N} or by a
relativistic jet \citep[see, e.g.,][]{2000A&A...362..113F} have both
been used to explain the spectra and inferred image sizes of \sgra.
Such conceptually different geometries will be readily distinguishable
using the combination of horizon-scale resolution, sensitivity to
different polarizations, and ability to follow the variability of
emission that will become available with the complete EHT array.

Our aim is to investigate the ability of future EHT observations to
distinguish between different emission geometries, black-hole
properties, and thermodynamic conditions in the accreting plasma
around \sgra.
In this first paper, we explore a large suite of GRMHD simulations
onto black-holes \citep{2012MNRAS.426.3241N, 2013MNRAS.436.3856S} with
different black-hole spins, different prescriptions regarding the
thermodynamic properties of the electrons, and different magnetic
field topologies of the saturated state of the turbulent flow.
Among the very large range of possible configurations, we select those
that agree with the broadband spectrum of \sgra\ as well as with the
initial measurement of the size of its image at 1.3\,mm.
In follow up articles, we will use this suite of models to make
detailed predictions for EHT observations and to develop observing
strategies that will maximize the scientific return of the EHT.

In earlier work, GRMHD simulations have been used to explore the
effect of changing the black-hole spin magnitude and orientation with
respect to the observer \citep{2010ApJ...717.1092D,
  2009ApJ...706..497M}, the tilt of the black-hole spin with respect
to the angular momentum of the accreting flow
\citep{2013MNRAS.432.2252D}, and the thermodynamic properties of the
electrons in the outflow/jet \citep{2013A&A...559L...3M}.
Our work improves on three aspects of these earlier studies.
First, we consider the influence of the large-scale magnetic field in
the saturated state of the flow, by contrasting the simulations with
Magnetically Arrested Disks (MAD) to those with Standard And Normal
Evolution (SANE) of \citet{2012MNRAS.426.3241N} and
\cite{2013MNRAS.436.3856S}.
These two types of simulations have very different dynamical
behaviors, as well as very different magnetic field topologies near
the black-hole horizon.
Second, the GRMHD simulations that we are using have been evolved long
enough for the flow to reach a dynamical steady state out to about a
hundred gravitational radii.
This is important because a very large volume of the accretion flow
contributes to the low-frequency synchrotron emission as well as to
the X-ray bremsstrahlung emission.
Finally, owing to our use of a very fast radiative transport
algorithm, instead of calculating the time-averaged properties of the
simulated flows before constructing images and spectra, we calculate
images and spectra for each snapshot of the flow, before averaging
them together.
Albeit computationally very expensive, our approach mimics more
closely the averaging that will inevitably occur during the EHT
observations and produces results that differ considerably from the
alternate procedure.

In order to achieve the high efficiency of ray tracing required to
calculate images and broadband spectra of every snapshot of a GRMHD
simulation, we employ our algorithm
\GRay\ \citep{2013ApJ...777...13C}.
Unlike standard central processing unit (CPU)-based ray tracing
algorithms, \GRay\ uses graphics processing units (GPUs) to accelerate
the computationally intensive geodesic integration.
With careful handling of the access to the memory, \GRay\ achieves an
order of magnitude speed up and allows us to compute
$\mathcal{O}(10^6)$ images during 12~hours of wall time using 32
nVidia Tesla K20X GPUs on the \texttt{El~Gato} cluster at the
University of Arizona.

In the next section, we describe the GRMHD simulations, the plasma
methods, our implementation of the radiative processes, and the ray
tracing algorithm that we are using in this paper.
In \S\ref{sec:properties}, we study the dependence of the calculated
spectra and images on the various model parameters.
In \S\ref{sec:size}, we compile the current spectral and imaging
observations of \sgra, which we then use in \S\ref{sec:fit} in order
to constrain the range of model parameters.
Finally, we discuss the implications of our results in
\S\ref{sec:discussions}.

\section{Accretion Models}

\begin{deluxetable}{lcccc}
  \tablewidth{\columnwidth}
  \tablecaption{Summary of the six sets of GRMHD models of black-hole
    accretion systems used in this study; their detailed descriptions
    can be found in \citet{2012MNRAS.426.3241N} and
    \citet{2013MNRAS.436.3856S}.}
  \tablehead{Model & \!\!\!\!\!\!Black Hole\!\!\!\!\!\! & Initial   & Resolution        & Snapshots Used \\
                 & Spin $a$                           & $B$ Field & $(r,\phi,\theta)$ & in $GM/c^3$}
\startdata
  \texttt{a0SANE}& 0.0 & multi-loop & $256\times128\times64$ &230,000--230,990\\
  \texttt{a7SANE}& 0.7 & multi-loop & $256\times128\times64$ &103,000--103,990\\
  \texttt{a9SANE}& 0.9 & multi-loop & $256\times128\times64$ &\ \,54,000--\ \,54,990\\
  \texttt{a0MAD} & 0.0 & single-loop& $264\times126\times60$ &210,000--210,990\\
  \texttt{a7MAD} & 0.7 & single-loop& $264\times126\times60$ &\ \,91,000--\ \,91,990\\
  \texttt{a9MAD} & 0.9 & single-loop& $264\times126\times60$ &\ \,47,000--\ \,47,990
\enddata

  \label{tab:sim}
\end{deluxetable}

\begin{figure*}
  \includegraphics[width=\textwidth,trim=0 0 0 0]{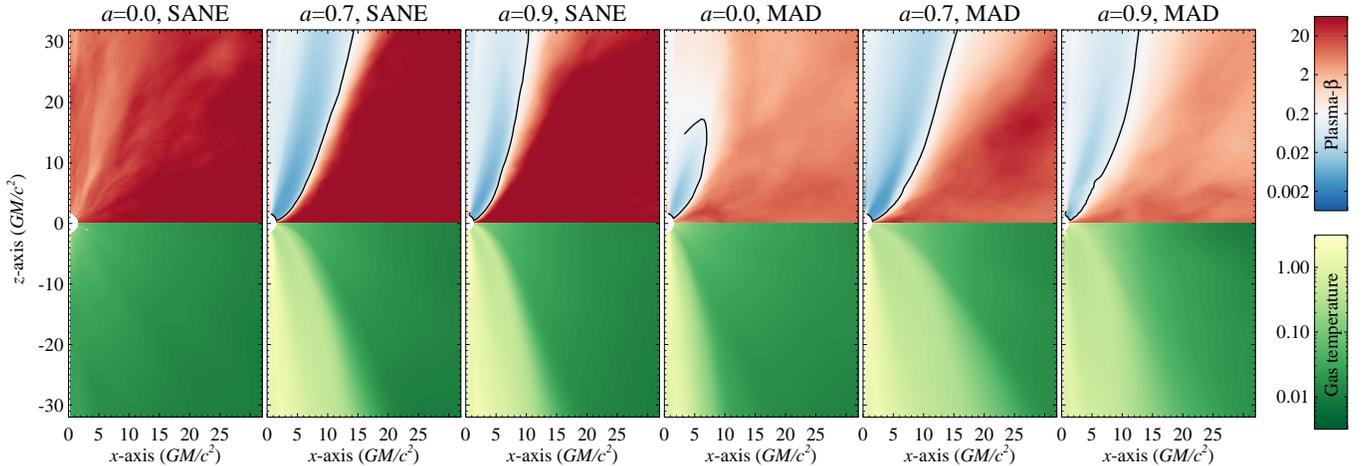}
  \caption{\emph{(Top)} Temporal and azimuthal averages of the
    plasma-$\beta$ for the suite of GRMHD simulation listed in
    Table~\ref{tab:sim}.
    The color scales for all the simulations are the same and shown in
    the right-most color bar.
    Blue represents $\llangle\beta\rrangle < 0.2$ and shows the
    strongly magnetized regions.
    Red represents $\llangle\beta\rrangle > 0.2$ and shows the gas
    dominated regions.
    The black solid lines mark the $\beta = 0.2$ contours.
    The saturated red regions around the equator of all the SANE
    simulation indicate that they are gas dominated, while the MAD
    simulations have stronger magnetic fields.
    \emph{(Bottom)} Temporal and azimuthal averages of the comoving
    dimensionless gas temperature $\llangle T\rrangle$.
    The color scales for all the panels are the same and shown in the
    right-most color bar.
    Green and yellow are low and high temperatures, respectively.
    Note that all simulations show a low plasma-$\beta$/high
    temperature funnel region except for the non-spinning SANE
    simulation \texttt{a0SANE}.
    The simulations are described in detail in
    \citet{2012MNRAS.426.3241N} and \citet{2013MNRAS.436.3856S}.}
  \label{fig:profiles}
  \vspace{6pt}
\end{figure*}

We carry out ray tracing calculations on six sets of three-dimensional
GRMHD simulation reported in \citet{2012MNRAS.426.3241N} and
\citet{2013MNRAS.436.3856S}, which consist of two different classes of
accretion flows.
The first class, called Standard And Normal Evolution (SANE), uses
multi-loop initial magnetic fields, while the second class, called
Magnetically Arrested Disk (MAD), uses single-loop initial fields.
The accretion flows were evolved for exceptionally long times, e.g.,
$\sim 200,000 GM/c^3$, until the flows reached steady state up to
$\sim 100GM/c^2$.
Three values of spin parameter, $a = 0$, 0.7, and 0.9 were used in
each class.
We summarize the setup of these simulations in Table~\ref{tab:sim}.

%------------------------------------------------------------------------------
\subsection{Characteristics of the MHD Simulations}
\label{sec:sim}

The differences in the initial magnetic fields and black-hole spins
affect the dynamic and thermodynamic properties of the accretion
flows.
Because the plasma-$\beta$ and the gas temperature are important
quantities in the emission models and affect the observables, we first
calculate, for each simulation, their temporal and azimuthal averages,
which we denote by
\begin{equation}
  \llangle\beta\rrangle \equiv
  \llangle P_\mathrm{gas} / P_\mathrm{mag}\rrangle,
\end{equation}
and
\begin{equation}
  \llangle T\rrangle \equiv \llangle u/\rho\rrangle\;,
  \label{eq:tdefn}
\end{equation}
and plot them in Figure~\ref{fig:profiles}.
In the above definitions, $P_\mathrm{gas}$ and $P_\mathrm{mag}$ denote
gas and magnetic pressure, while $u$ and $\rho$ denote the internal
energy and density.

All simulations show a funnel region with low plasma-$\beta$ and high
temperature, with only exception the zero spin SANE simulation
\texttt{a0SANE}.
This makes \texttt{a0SANE} a special configuration in our subsequent
studies.
The transition from the funnel region to the disk occurs over a rather
narrow sheath, across which the plasma-$\beta$ changes rapidly (note
the logarithmic scale for the colors).
This is consistent with the findings of \citet{2013A&A...559L...3M}
and allow us to separate the funnel from the disk by setting a
threshold on the plasma-$\beta$, which here we fix to $\Bt=0.2$ (see
\S\ref{sec:fit}).
The saturated red region around the equators of all the SANE
simulations indicates that the accretion disks are strongly gas
dominated.
The MAD simulations, on the other hand, tend to have stronger magnetic
fields even in the disk regions.
For more details of the simulations, we refer to the original papers
by \citet{2012MNRAS.426.3241N} and \citet{2013MNRAS.436.3856S}.

We note that the coordinate singularities along the poles can cause
numerical difficulties and can lead to a few low-density hot cells.
Although these artificial hot cells do not affect the flow dynamics,
they may change the predicted spectra by over producing X-rays and
$\gamma$-rays.
In addition, the simulations use a special coordinate transformation
developed by \cite{2011MNRAS.418L..79T} to ``cylindrificate'' the grid
near the poles.
While this technique significantly speeds up the GRMHD simulations by
allowing a larger time step, it expands the problematic hot cells and
enlarges the error.
To overcome these numerical artifacts, we excise the cells around the
poles by setting their emissivity to zero when they are unphysically
hot compared to their neighbors.

The six numerical models described above are scale free in length and
density (their time, velocity, and energy scales can be obtained by
scaling length and density with different powers of the speed of light
$c$).
In order to compute images for \sgra, we fix the length scale by
setting the mass of the central black-hole to that of \sgra,
$M_\mathrm{bh} = 4.3\times10^6 M_\odot$ \citep{2009ApJ...692.1075G}.
We introduce the density scale $\Ne$ as a free parameter to convert
the gas density $\rho$ into the electron number density, which at the
same time determines the physical accretion rate.
We also define the observer's inclination angle $i$ with respect to
the spin axis of the black-hole and treat it as a free parameter.

%------------------------------------------------------------------------------
\subsection{Thermodynamics of the Accretion Flow}
\label{sec:sim}

To calculate spectra, lightcurves, and images of the accretion flow
around \sgra, we need to specify the electron temperature in the disk
and in the funnel regions, which depend on the details of the heating
due to the dissipation of turbulence, the radiative cooling, as well
as the energy exchange between the protons and the electrons in the
flow.
In low density regions, where the collisional timescale is long
compared to the dynamical timescale, the electron temperature $\Te$ is
expected to be lower than the ion temperature $\Ti$, leading to a
two-temperature plasma (\citealt{1976ApJ...204..187S,
  1995Natur.374..623N, 1998ApJ...492..554N}; see also the recent
review by \citealt{2014ARA&A..52..529Y} and references therein).
In addition, in the funnel region, where the heating rate due to the
dissipation of the MHD turbulence is expected to be low, thermal
conduction can be efficient in bringing the electrons to a constant
temperature \citep{2013A&A...559L...3M}.
We treat both of these possibilities in the calculation of the
radiation from the accretion flow and use the following
parametrizations to capture the resulting electron temperatures.

\emph{Constant Electron-Ion Temperature Ratio Model.---\/}In the first
model for the plasma, we assume that the electron-to-ion temperature
ratio is a fixed function of the plasma-$\beta$.
This is motivated by the fact that the temperature of the electrons
can differ significantly from that of the protons as the gas density
gets lower and the magnetic field strength, which influences the
cooling time for the electrons via synchrotron emission, gets higher.
We consider a simple step-function model where the electron
temperature takes the following form:
\begin{equation}
  \Te / \Ti = \begin{cases}
    \td & \mbox{if } \beta > \Bt, \\
    \tf & \mbox{otherwise.}
  \end{cases}
\end{equation}
In this equation, the symbols $\td$ and $\tf$ denote the electron
temperatures in the disk and in the funnel, in units of the ion
temperature.
Both quantities are expected to be less than unity under the
assumptions that \emph{(i)} dissipation of turbulence mainly heats up
the ions, \emph{(ii)} the electrons cool faster than the ions, and
\emph{(iii)} the electron cooling rate is faster than the energy
exchange rate between electrons and ions \citep[see discussion
  in][]{1994ApJ...428L..13N, 1995ApJ...452..710N, 1999ApJ...520..248Q}.
We specify the electron-to-ion temperature ratio in the disk, $\td$,
and the electron-to-ion temperature ratio in the funnel, $\tf$, and
constrain these parameters in the next section using observations of
\sgra.
Note that this model, which we will refer to as the ``constant ratio
model'' in the remainder of the paper, reduces to the standard
two-temperature plasma model used in earlier studies \citep[e.g.,][]{
  2009ApJ...703L.142D, 2009ApJ...706..497M}, if we fix $\td = \tf$.

\emph{Constant Funnel Electron Temperature Model.---\/}The second
model accounts for the possible effects of electron conduction in the
funnel region.
We follow the parametrization in \citet{2013A&A...559L...3M} and
assume that the electron temperature in the funnel is constant.
We retain the constant temperature ratio parametrization for the disk.
Because the plasma-$\beta$ is a good indicator of the separation
between the highly-magnetized, low density funnel region and the
higher density disk, we use a threshold value of this parameter to
specify the change in the electron temperature, such that
\begin{equation}
  \Te = \begin{cases}
    \Ti \td & \mbox{if } \beta > \Bt, \\
    \Tf     & \mbox{otherwise.}
  \end{cases}
\end{equation}
Note that $\Tf$ may acquire values larger than unity for
ultrarelativistic electrons according to equation~(\ref{eq:tdefn}).
We vary the values of the threshold plasma-$\beta$, $\Bt$, the
electron-to-proton temperature ratio in the disk, $\td$, and the
electron temperature in the funnel, $\Tf$, to investigate the effects
of these three parameters on the observables and when fitting the
observed spectrum and images of \sgra.
We refer to this as the ``constant temperature model'' in the rest of
the paper.

%------------------------------------------------------------------------------
\subsection{Radiative Processes and Transfer}
\label{sec:jnu}

We calculate the radiation emitted from the accretion flow by solving
the radiative transfer equation along null-geodesics through the
domain of the GRMHD simulations, using the \GRay\ code.
\GRay\ integrates the radiative equation backward from the image plane
to the source.
This approach has at least two advantages.
First, it allows us to solve only for light rays that are normal to
the image plane, drastically reducing the number of photon
trajectories that need to be integrated.
Second, integrating the radiative transfer equation backward allows us
to stop the integration when the optical depth is sufficiently large
(we use a cutoff at $\ln 1000 \approx 6.9$) or when the outgoing ray
is sufficiently far away ($\sim 1000 GM/c^2$) from the
black-hole\footnote{The actual criterion is available in the
  \GRay\ source code, \url{https://github.com/chanchikwan/gray}.
  All results presented in this paper should be reproducible (within
  round-off error) by using commit \texttt{0c99a24c} and
  \texttt{CUDA~6.0.1}.}.
This significantly speeds up the image calculation for optically thick
media.
Both of these advantages make ray tracing algorithms much faster than
Monte Carlo techniques for our application.

To ensure numerical stability, we follow \citet{2012A&A...545A..13Y}
to express the radiative transfer equation in two coupled differential
equations:
\begin{align}
  \frac{d\tau}{d\lambda}
  &= \gamma^{-1} \alpha_{0,\nu},
  \nonumber\\
  \frac{d\mathcal{I}}{d\lambda}
  &= \gamma^{-1} \left(\frac{j_{0,\nu}}{\nu^3}\right) e^{-\tau},
  \nonumber
\end{align}
where $\lambda$ is the affine parameter, $\gamma^{-1} \equiv
\nu_0/\nu$ is the relative energy shift, $\tau$ and $\mathcal{I}$ are
the optical depth and Lorentz invariant intensity at frequency $\nu$,
and $\alpha_{0,\nu}$ and $j_{0,\nu}$ are comoving absorption and
emission coefficients.

\begin{figure*}
  \includegraphics[width=\textwidth,trim=0 0 0 9]{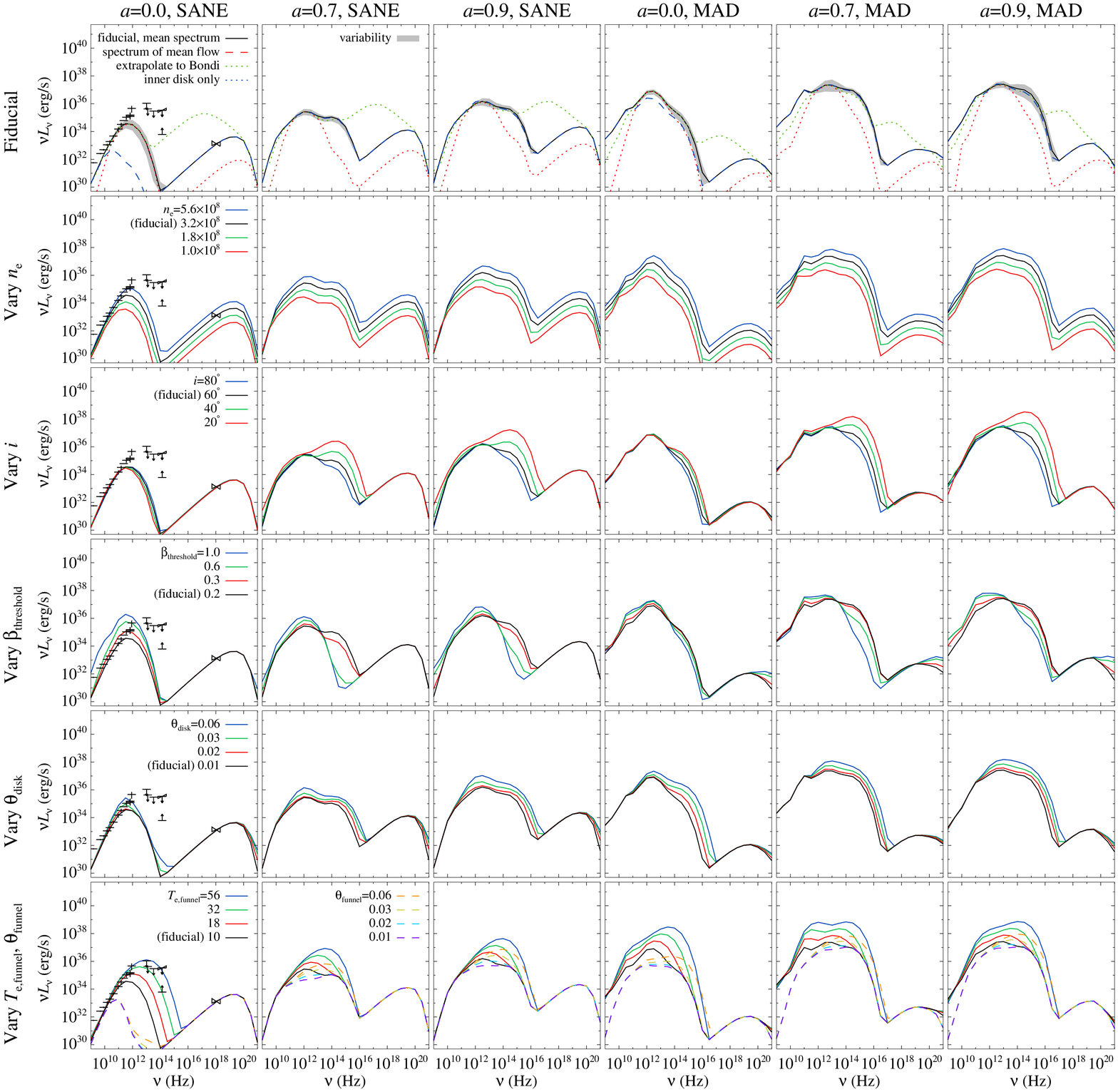}
  \caption{Broadband spectra computed for the suite of GRMHD
    simulations listed in Table~1.
    Each column corresponds to a different simulation, while each row
    shows the effect of varying one of the model parameters.
    The data points in the first column correspond to various
    observations of \sgra.
    The top ``Fiducial'' row uses parameters $i = 60^\circ$, $\Ne =
    3.2\times10^{8}$, $\Bt = 0.2$, $\td = 0.01$, and $\Tf = 10$.
    In each panel, the shaded gray area marks the variability of
    spectra computed from 100 snapshots; the black solid curve marks
    their average; the red dotted curve marks the average spectrum of
    the inner accretion flow; the green dotted curve marks the average
    spectrum computed after extrapolating the properties of the flow
    out to the Bondi radius; and the blue dashed curve marks the
    spectrum of the mean flow.
    We vary the normalization of the electron density $\Ne$ in the
    second row, the observer's inclination $i$ in the third row, the
    threshold plasma-$\beta$ in the fourth row, and the electron-ion
    temperature ratio $\td$ in the disk in the fifth row.
    In the sixth row, we either vary the funnel electron temperature
    $\Tf$ (solid curves) or the funnel electron-ion temperature ratio
    $\tf$ (dashed curves).}
  \label{fig:var}
  \vspace{6pt}
\end{figure*}

In hot, magnetized accretion flows, synchrotron and bremsstrahlung
radiation are the two major radiative processes that contribute to the
emission and absorption \citep[see][]{1998ApJ...492..554N}.
In this paper, we assume thermal synchrotron and bremsstrahlung
emission and do not treat the possible contribution from non-thermal
electrons, which can contribute to both radio and X-ray fluxes
\citep{1998Natur.394..651M, 2000ApJ...541..234O} and can help explain
the variability observed in the X-rays \citep{2009ApJ...701..521C}.
For synchrotron emissivities, we use the approximate expression
derived by \citet{2011ApJ...737...21L}.
For thermal bremsstrahlung emission, we use the expression derived in
\citet{1979rpa..book.....R}, with a Gaunt factor taken from
\citet{1973blho.conf..343N}\footnote{Although
  \citet{1979rpa..book.....R} cited \citet{1973blho.conf..343N} for
  their Gaunt factor, the actual formulae are different.
  See the \GRay\ source code for our implementation of the Gaunt
  factor.}.
Although Compton scattering is generally important in stellar-mass
black holes, its contribution to the \sgra\ spectrum is mainly in the
optical \citep[see, e.g.,][]{1998ApJ...492..554N}, for which we do not
have data to impose any constraints on (see \S\ref{sec:size}).
Moreover, \citet[see their Figure~4]{2009ApJ...706..497M} showed that
the contribution of Compton scattering to the X-ray flux predicted in
GRMHD simulations is typically much smaller than the observed flux,
unless the black hole is rapidly spinning and observed from a nearly
edge-on orientation.
For these two reasons, we neglect here the effects of Compton
scattering on the spectrum.

\section{Spectral and Image Properties}
\label{sec:properties}

\begin{figure*}
  \includegraphics[width=\textwidth,trim=0 0 0 0]{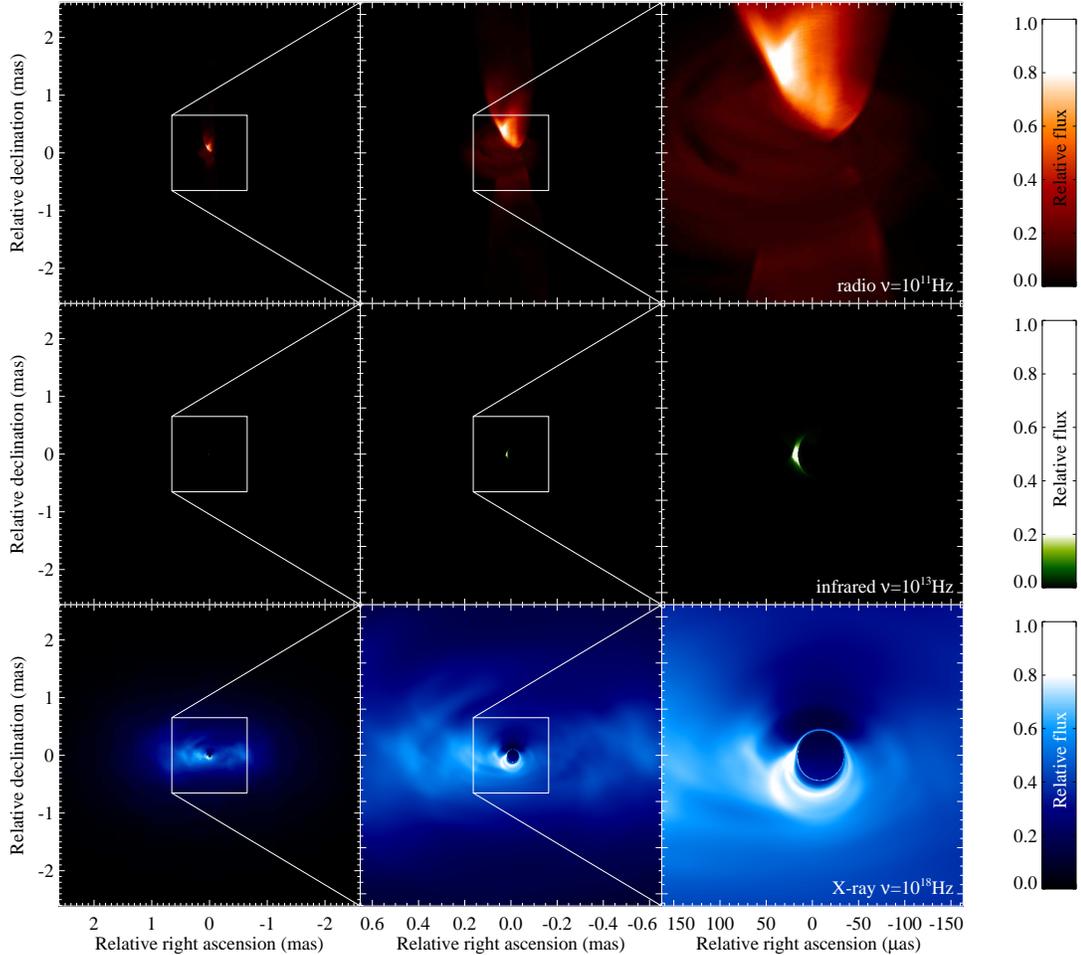}
  \caption{Simulated images of \sgra\ at three frequencies using the
    simulation \texttt{a9MAD} with the fiducial model parameters.
    The color scale for each row is shown on the right, with unity
    being the maximum flux (the same color scales are used in the
    subsequent image plots).
    The different columns, from left to right, depict images computed
    on different zoom scales of $512\rS\times512\rS$,
    $128\rS\times128\rS$, and $32\rS\times32\rS$.
    In order to preserve individual visible features in the accretion
    flows, we average only the ten last snapshots, i.e., between
    $t=$47,900 and 47,990.
    The low frequency radio and the X-ray emission originate from
    large volumes, whereas the infrared emission primarily arises from
    the inner accretion flow.}
  \label{fig:zoom}
  \vspace{6pt}
\end{figure*}

Using \GRay\ for radiation transport and the plasma models described
in the previous section, we calculate the broadband spectra emerging
from the accretion flow in each GRMHD simulation.
We compute a radiation spectrum from a series of snapshots taken from
the GRMHD simulations.
In order to take into account the emission from near the event horizon
as well as at the outer parts of the computational domain, we compute
the images using three different fields of view: the innermost images
cover a $32\rS\times32\rS$ plane, the middle images cover a
$128\rS\times128\rS$ plane, and the outermost images cover a
$512\rS\times512\rS$ plane, where $\rS \equiv 2GM_\mathrm{bh}/c^2$ is
the Schwarzschild radius.
The outermost image size is chosen so that the X-ray fluxes computed
within the simulation volume converge.
We resolve each domain using $512 \times 512$ pixels and obtain the
total flux by appropriately combining and summing the contributions
from the three images.

In order to explore the effect of the parameters describing the
thermodynamics of the plasma, as well as of the inclination of the
observer with respect to the spin axis of the accretion flow, on the
spectra and images of the flows, we first calculated a large suite of
simulations for a range of values of the parameters.
Each of the two plasma models described in the previous section
requires five parameters to be fully specified.
For the constant ratio model, these are: the normalization of the
electron density $\Ne$, the inclination of the observer $i$, the
threshold plasma-$\beta$, $\Bt$, that separates the funnel from the
disk, the electron-to-ion temperature ratio in the funnel, $\tf$, and
the electron-to-in temperature ratio in the disk, $\td$.
For the constant temperature model, these are: the normalization of
the electron density $\Ne$, the inclination of the observer $i$, the
threshold plasma-$\beta$, $\Bt$, the funnel electron temperature
$\Tf$, and the electron-to-ion temperature ratio, $\td$ in the disk.

In Figure~\ref{fig:var}, we show the effect of independently varying
each of these parameters on the simulated radio-to-X-ray spectrum for
\sgra.
Each column corresponds to one of the GRMHD simulations listed in
Table~1, while each row shows how the spectra change in response to
varying one of the five model parameters, while keeping the others
fixed.
The observational data points shown in the leftmost panels were
collected by \citet{2011ApJ...735..110B}.

As mentioned at the beginning of this section, the X-ray fluxes
computed within the simulation volume converge at a radius of $\sim
512\rS$.
However, this radius is much smaller than the Bondi radius of \sgra,
$\sim 10^5\rS$, at which the accretion flows is expected to still have
significant contribution to the X-ray emission \citep[see, e.g.,][]{
  2004ApJ...613..322Q}.
Our models do not include this large volume of X-ray emission because
the central black-hole is fed by a torus that lies at a few hundred
Schwarzschild radii.
To consider the contribution of the large scale flow, we computed a
set of spectra up to the Bondi radius by using the extrapolation
scheme developed in \citet{2013MNRAS.432..478S}.
The results are shown as green dotted curves in the first row in
Figure~\ref{fig:var}.
The X-ray fluxes of the extrapolated flows are one-to-two orders of
magnitude larger, as expected.
Note that even though we show the results from this extrapolation
here, we will opt to limit our domain to the inner accretion flow and
consider the fraction of X-rays that originate from the inner flow
when actually fitting the X-ray data to the models (see Section 4).

\emph{Fiducial model.---\/}The resulting broadband spectrum for each
GRMHD simulation and a ``fiducial'' plasma model with a representative
set of parameters ($\Ne = 3.2\times10^{8}$, $i = 60^\circ$, $\Bt =
0.2$, $\td = 0.01$, and $\Tf = 10$) is depicted in the first row of
Figure~\ref{fig:var}.
Note that this model is not a specific fit to \sgra\ observations.
We use the fiducial setup, before embarking on fitting the
observations, in order to study several theoretical aspects of the
spectra, such as the predicted variability across the spectrum and the
different emission regions that give rise to the structures of the
images at different wavelengths.

The strong turbulence in the accretion flow naturally causes the
resulting spectra and images to be time dependent.
This complicates the comparison of the simulations to the
observations, as both need to be averaged properly in order to avoid
comparing a particular realization of the turbulent flow in the
simulations to a different realization in the observed flow.
In order to assess the variability of the simulated spectra and images
we perform our ray tracing calculation for 100 different snapshots in
each of the simulations, as listed in Table~\ref{tab:sim}.
Each snapshot of the simulation was obtained at regular time intervals
of $10 GM/c^3$.
For the mass of \sgra, 100 snapshots correspond to $\approx
5.9$~hours, which is similar to the time interval over which EHT
observations will take place.

The black solid curves on the panels of the first row in
Figure~\ref{fig:var} show the mean spectra obtained by averaging the
simulated spectra in the 100 snapshots of each simulation.
The shaded gray area around them marks the maxima and minima of the
spectra emerging from these 100 snapshots and is representative of the
expected spectral variability.
The red dotted curves on the same panels show the average spectra
computed using only the image of size $32\rS\times32\rS$, which we
refer to as the emission from the inner accretion flow.
As expected, when the emission originates from the inner accretion
flow, the spectra acquire their largest degree of variability while
the opposite is true when the emission originates in a much larger
volume.
Indeed, in the optically thick, low-frequency region of the spectrum,
the emission is weakly variable because the radius of the photosphere
at these frequencies is equal to tens to hundreds of Schwarzschild
radii and the dynamical timescales there are very long.
The variability is also weak at the optically thin, high-frequency
region of the spectrum, where the emission is generated by thermal
bremsstrahlung over a very large volume around the black-hole.
In the millimeter to IR range (i.e., $\nu \sim
10^{11}$--$10^{14}\,\mathrm{Hz}$ for the \texttt{a0SANE} simulation),
however, the emission is optically thin and originates very close to
the event horizon.
The characteristic timescales there are very short and both the
spectra and images show significant variability.

To visualize this point in a different manner, we show in
Figure~\ref{fig:zoom} the averaged snapshot images of the
\texttt{a9MAD} simulation, for the fiducial parameters of the plasma
model.
The different rows, from top to bottom, depict images at frequencies
$\nu = 10^{11}\,\mathrm{Hz}$, $10^{13}\,\mathrm{Hz}$, and
$10^{18}\,\mathrm{Hz}$.
The different columns, from left to right, are for image sizes equal
to $512\rS\times512\rS$, $128\rS\times128\rS$, and $32\rS\times32\rS$.
As discussed above, the emission at the lowest and highest frequencies
shown originates primarily from large distances away from the
black-hole.
On the other hand, the emission at infrared frequencies originates in
a very small region, close to the horizon.
The black-hole shadow is obscured at the optically thick radio
frequencies but is visible in both infrared and X-ray, which are
optically thin.
For the X-ray images, although the region around the black-hole shadow
has a larger surface brightness than all other regions, most of the
(integrated) flux actually originates from a few hundred Schwarzschild
radii in our models.

In calculating the average spectra and images discussed above, we
computed individual spectra and images for each snapshot of the
simulations and then averaged together the resulting surface
brightness and fluxes on the image plane of the observer.
This procedure generates results that can be very different compared
to calculating the average hydrodynamic and thermodynamic properties
of each simulation and then computing a single image and spectrum for
this mean flow.
This is because of the fact that the plasma properties are
substantially variable and the radiative transfer equation, which we
solve along geodesics to calculate the image brightness, is a highly
non-linear function of the plasma properties.
In the top row of Figure~\ref{fig:var}, we show as blue dashed curves
the spectra computed using the mean properties of the flows.
The relative difference between the two averaging procedures is
largest in the case of the non-spinning black-holes.

\begin{figure}
  \includegraphics[width=\columnwidth,trim=0 0 0 0]{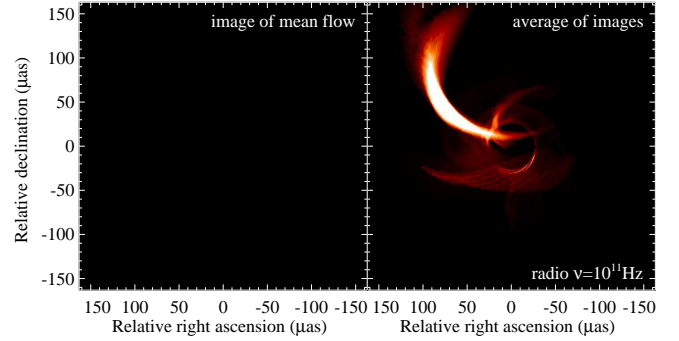}
  \caption{A comparison of the predicted image at $\nu =
    10^{11}\,\mathrm{Hz}$ using the average properties of the
    simulation (left) to the average of individual images computed for
    each snapshot of the simulation (right).
    For this figure, we used the \texttt{a0SANE} simulation with the
    parameters of the fiducial model.
    The striking difference between the images is due to the presence
    of short-lived, magnetically-dominated filaments in the inner
    accretion flow, which contribute significantly to the emission in
    each snapshot but are washed out when computing the mean flow.}
  \label{fig:comp}
  \vspace{6pt}
\end{figure}

\begin{figure*}
  \includegraphics[width=\textwidth,trim=0 0 0 3]{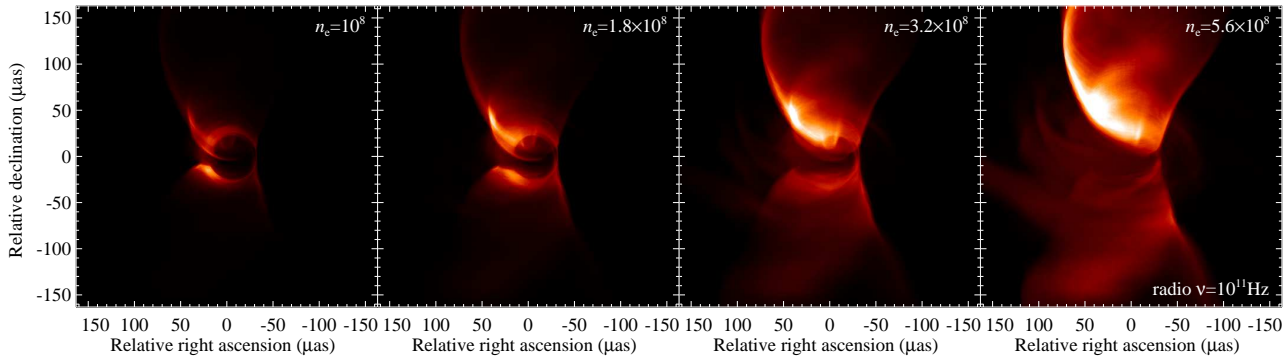}
  \caption{The effect of varying the electron density normalization
    $\Ne$ on the radio image at $\nu = 10^{11}\,\mathrm{Hz}$ for the
    \texttt{a7SANE} simulation.
    The lines of sight through the funnel (regions above and below the
    origins) are always optically thin at this frequency for this
    range of densities and their brightness scales as $\Ne^2$ (see text).
    In contrast, as the electron density increases, a larger part of
    the colder accretion disk (regions left and right of the origins)
    becomes optically thick, obscuring an increasing fraction of the funnel.
    The net effect is a dependence of the overall flux on electron
    density scale that is weaker than $\Ne^2$ at this frequency.}
  \label{fig:varne}
  \vspace{6pt}
\end{figure*}

In Figure~\ref{fig:comp}, we compare the images computed by the two
averaging approaches for a frequency of $10^{11}\,\mathrm{Hz}$, using
the \texttt{a0SANE} simulation.
The left panel shows the image calculated using the mean properties of
the flow and the right panel shows the average of the images in each
snapshot.
There is a striking difference between the two images.
The image of the mean flow (left panel) is very dim and almost
invisible in the plot.
On the other hand, the average image of the snapshots is dominated by
magnetic filaments, which are short lived (and hence do not contribute
significantly to the average properties of the flow) but are very
bright (and hence dominate the average emission).

Finally, comparing the spectra calculated for the different GRMHD
simulations but for the same, fiducial plasma model, we find that the
flux at the thermal peak increases monotonically as we move from the
leftmost to the rightmost columns.
In other words, there is more low-frequency emission in the MAD and in
the high-spin simulations.
This is a direct consequence of the fact that MAD and high-spin
simulations are characterized by significant relativistic jets.
In contrast, the MAD simulations generate less X-ray radiation than
the SANE simulations because there is less flux from the accretion
flows at large radii (i.e., $r \gtrsim 128\rS$).
This is also a direct consequence of the fact that the density in the
MAD simulations is more centrally concentrated than in the SANE
simulations \citep[see][]{2012MNRAS.426.3241N}.

\emph{Parameter study.---\/}The remaining rows in Figure~\ref{fig:var}
show how the spectra are affected when we vary one of the parameters
of the plasma model, i.e., $\Ne$, $i$, $\Bt$, $\Tf$, $\td$, and $\tf$,
while holding all other parameters fixed to the fiducial model.

We first vary the density scale in the range $\Ne = 10^8$ to
$\Ne=5.6\times10^8$ and show the result for each GRMHD simulation in
the second row of Figure~\ref{fig:var}.
At frequencies $\nu \gtrsim 10^{12}\,\mathrm{Hz}$, increasing the
density scales causes an increase in the overall flux that is
proportional to $\Ne^2$.
This dependence is trivial to understand for the X-rays, which are
generated by optically thin bremsstrahlung emission, since the
emissivity of bremsstrahlung scales as the square of the electron
density.
In the infrared, near the peak of the thermal bump, the emission is
primarily due to optically thin synchrotron processes, with an
emissivity that scales as $j_\mathrm{synch} \propto \Ne B^2$.
However, the natural scaling of the GRMHD equations causes the
dimensional magnetic field strength in the flow to be always
proportional to the square root of the electron density.
In other words, $B \propto \sqrt{\Ne}$, and the optically thin
synchrotron emissivity also scales as $j_{\rm synch}\propto \Ne^2$.

At lower frequencies, i.e., when $\nu \lesssim 10^{12}\,\mathrm{Hz}$,
the emerging flux has a weaker dependence on the density scale.
In this part of the spectrum, the inner accretion flow becomes
optically thick to synchrotron self absorption.
Increasing the density scale causes the cooler outer disk to partially
block the hotter inner disk and, therefore, to flatten the dependence
of the flux on the electron density.
To demonstrate this point, we show in Figure~\ref{fig:varne} the
images at a frequency of $\nu=10^{11}\,\mathrm{Hz}$ for the
\texttt{a7SANE} simulation and for the four different values of the
electron density scale.
As we increase the density scale, the inner accretion disk and the
funnel become brighter as expected.
An increasing fraction of that bright emission, however, is obscured
by the colder outer disk, reducing the strong dependence of the flux
on the scale of the electron density.

\begin{figure*}
  \includegraphics[width=\textwidth,trim=0 0 0 0]{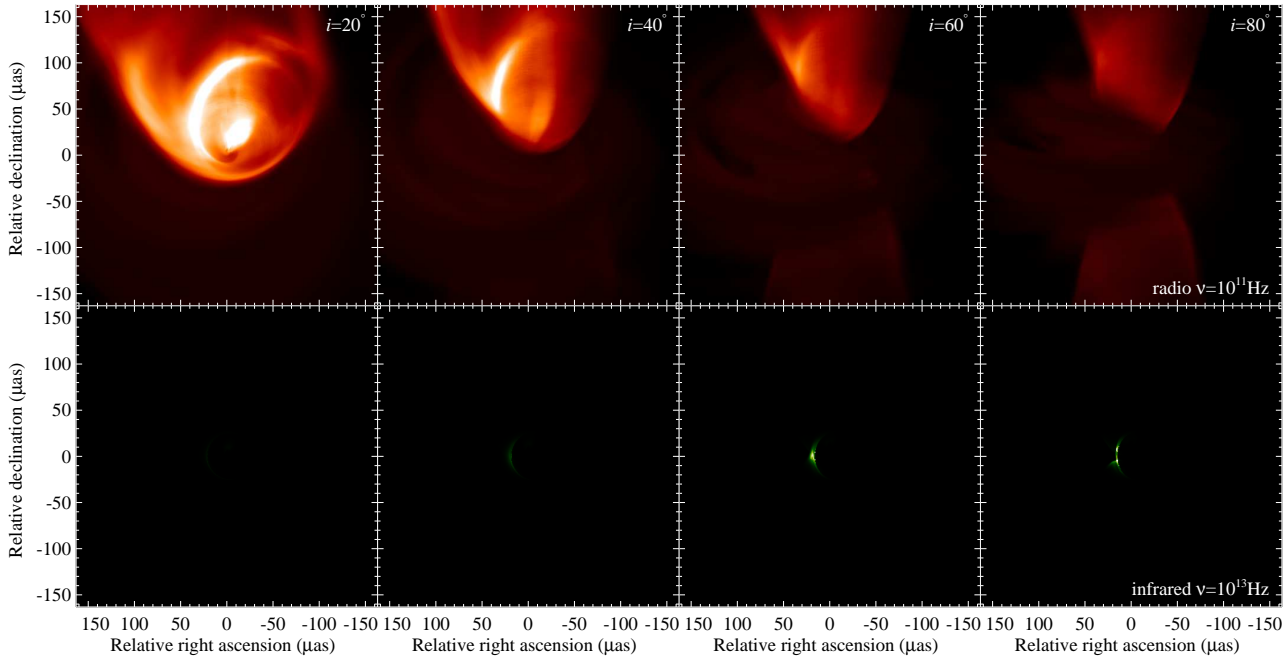}
  \caption{The effect of varying the observer's inclination on the
    predicted images for the \texttt{a9SANE} simulation at two
    different frequencies.
    At low frequencies, e.g., at $\nu = 10^{11}\,\mathrm{Hz}$ shown in
    the top row, as the observer's inclination increases, a larger
    fraction of the hot inner flow is obscured by the colder outer disk.
    This effect cancels out the fact that both the upper and lower
    funnel regions become visible, and produces an overall weak
    dependence of the inclination.
    In contrast, at intermediate frequencies, e.g., at $\nu =
    10^{13}\,\mathrm{Hz}$ shown in the bottom row, the optically thin
    accretion disk never obscures the funnel and the highest flux
    occurs at the highest inclinations, where Doppler effects are maximal.
    The observer's inclination has a negligible effect on the X-rays
    (not shown here), which originate from an optically thin,
    quasi-spherical large volume.}
  \label{fig:vari}
  \vspace{6pt}
\end{figure*}

In the third row of Figure~\ref{fig:var}, we show the effect of
varying the inclination of the observer in the range $i= 20^\circ$ to
$i=80^\circ$.
At low frequencies, i.e., when $\nu \lesssim 10^{12}\,\mathrm{Hz}$,
the flux has a very weak dependence on the inclination.
The dependence becomes stronger, however, at somewhat higher
frequencies.
In order to understand this behavior, we show in Figure~\ref{fig:vari}
the corresponding images of the accretion flow in the \texttt{a9SANE}
simulation.
At low frequencies (top row), as the observer's inclination increases,
a larger fraction of the hot inner flow is obscured by the colder
outer disk.
This effect cancels out the fact that both the upper and lower funnel
regions become visible, and produces an overall weak dependence of the
inclination.
At intermediate frequencies (bottom row), the optically thin emission
originates very close to the black-hole shadow; increasing the
inclination causes the Doppler effect to boost more radiation toward
the observer and, therefore, increases the resulting flux.
In X-rays (not shown in the figure), which originate in a large,
quasi-spherical volume, there is very little dependence of the
resulting spectrum on the inclination.

In the fourth row of Figure~\ref{fig:var}, we show the effect of
varying the threshold plasma-$\beta$ that distinguishes the disk from
the funnel, in the range $\Bt = 0.2$ to $\Bt= 1.0$.
For the parameters of the fiducial model, the funnel is brighter than
the disk in the radio to infrared frequencies.
Therefore, increasing $\Bt$, which creates a larger funnel, increases
the radio and infrared flux for all simulations.
The situation is reversed, however, at higher frequencies, where the
emission is dominated by the disk as shown in the lower row of
Figure~\ref{fig:vari}.
Note also that, primarily in the MAD simulations, the $\gamma$-ray
flux is also affected by the threshold value of the plasma-$\beta$.
This arises from the few hot zones along the pole near the event
horizon, as we described in \S\ref{sec:sim}, and we have simply left
it here to demonstrate this numerical artifact.

In the fifth row of Figure~\ref{fig:var}, we show the effect of
varying the electron-to-ion temperature ratio in the disk, in the
range $\td = 0.01$ to $\td=0.06$.
As expected, because the synchrotron and bremsstrahlung emissivities
depend on the electron temperature, the flux in the optically thin
part of the radio-to-infrared spectrum increases with $\td$.

Finally, in the sixth row of Figure~\ref{fig:var}, we show the effect
of varying the thermodynamic properties of the electrons in the
funnel.
In the case of the constant temperature model, we vary the electron
temperature in the funnel in the range $\Tf=10$ to $\Tf=56$; in the
case of the constant ratio model, we vary the electron-to-ion
temperature ratio in the funnel in the range $\tf=0.01$ to $\tf=0.06$.
As discussed above, for the fiducial values of the parameters, the
emission from the funnel dominates the radio-to-infrared part of the
emission.
As a result, changing the electron temperature in the funnel has a
very large effect on that part of the spectrum.
On the other hand, the X-ray emission comes primarily from the outer
disk and, therefore, is not affected by the thermodynamic properties
of the electrons in the funnel.

\section{Current Spectral and Imaging Observations of \sgra}
\label{sec:size}

\begin{figure*}
  \includegraphics[width=\textwidth,trim=0 0 0 0]{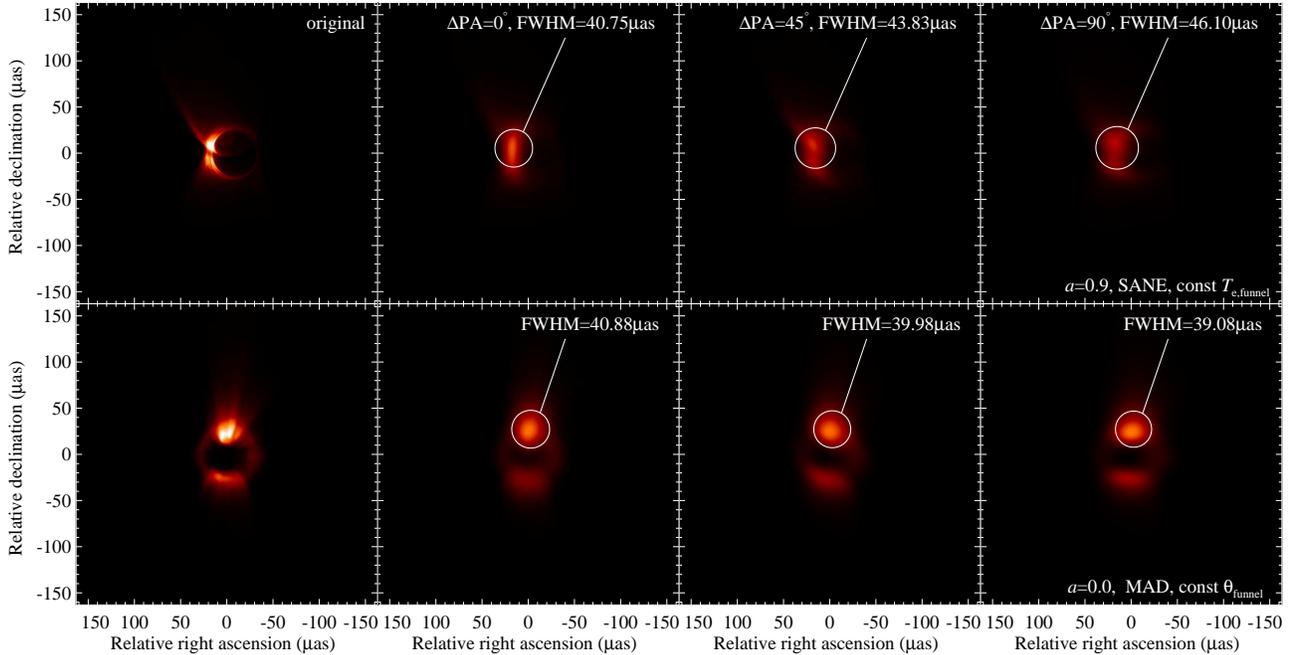}
  \caption{Calculated images at 1.3\,mm for \emph{(Top)} the
    \texttt{a0SANE} simulation using the constant ratio model and
    \emph{(Bottom)} the \texttt{a9SANE} simulation using the constant
    temperature model.
    All of the parameters of the plasma model correspond to the
    best-fit configurations in each case (and not to the fiducial
    model, as in the previous figures).
    The leftmost panels are the results of the direct ray tracing
    calculations.
    The remaining columns show the effect of broadening caused by
    interstellar scattering.
    Because the relative position angle, $\DPA$, between the spin axis
    of the black-hole and the major axis of the scattering kernel is
    not known, the three columns show the resulting images for three
    different relative position angles as examples.}
  \label{fig:dpa}
  \vspace{6pt}
\end{figure*}

In \S3, we performed an extensive parameter study aiming to understand
the dependence of our results on the model assumptions related to the
MHD and thermodynamic properties of the accretion flow.
Our next goal is to identify the set of models and parameters that are
consistent with the current spectral and imaging observations of
\sgra.
The data points in the leftmost panels of Figure~\ref{fig:var}
correspond to non-simultaneous observations of \sgra\ at a broad range
of frequencies and with different instruments.
Several different, partially overlapping data sets exist in the
literature.
The particular data points we are using here were prepared by
\citet{2011ApJ...735..110B}, while a comparable collection of data is
available in the more recent review by \citet{2013CQGra..30x4003F}.

When comparing the results of our simulations to spectral
observations, we make the following choices.\\
\indent\emph{(i)} In the radio, we do not consider the data points at
frequencies below $10^{11}$~Hz, because even a small fraction of
non-thermal electrons in the accretion flow (which we do not include
here) can affect significantly the flux at these low frequencies
\citep[see][]{1998Natur.394..651M, 2000ApJ...541..234O,
  2009ApJ...701..521C}.\\
\indent\emph{(ii)} In the infrared, i.e., at frequencies
$10^{13}$~Hz$<\nu<10^{15}$~Hz, we do not consider the individual data
points but rather require the simulated spectra to fall within the
lowest and highest observed fluxes at $\nu=1.38\times 10^{14}$~Hz.
This is justified by the fact that both the observational data
\citep{2003Natur.425..934G, 2004ApJ...601L.159G, 2011ApJ...728...37D}
and our simulations shown in Figure~\ref{fig:var} indicate that the
infrared emission is highly variable and specific observed or
simulated fluxes will depend entirely on the specific realization of
the turbulent flow.\\
\indent\emph{(iii)} In the X-rays, we investigated the option of
extrapolating the accretion flow out to the Bondi radius following the
prescription of \citet[see also discussion of
  Figure~\ref{fig:var}]{2013MNRAS.432..478S} and computing the X-ray
emission using this entire volume.
We found, however, that our results depended very strongly on the
assumed power-law indices of the density and temperature profiles as
well as on the choice of the point at which the extrapolation matched
the numerical solutions.
Instead of following this approach, we opt to use the result of
\citet[see also the discussion in
  \citealt{2013ApJ...774...42N}]{2010ApJ...716..504S} that 10\% of the
quiescent X-ray flux from Sgr A* originates in a point source and
attribute 10\% of the observed flux to the emission from our simulated
volume.

The size of the image of \sgra\ has been measured over many
wavelengths, from the radio to the millimeter (see
\citealt{2013CQGra..30x4003F} for a recent review).
At most wavelengths, the size measurement is dominated by the blurring
of the image caused by interstellar scattering.
There is strong evidence that, at wavelengths below 1\,cm, the
intrinsic size of \sgra\ can be discerned \citep[see,
  e.g.,][]{2014ApJ...790....1B}.
However, because of the $\lambda^2$ dependence of the size of the
scattering ellipse, the most accurate measurements occur at the
smallest wavelengths.
Early EHT observations of \sgra\ at 1.3\,mm measured its size at
$43^{+14}_{-8}\,\mathrm{\mu as}$.
Correcting for the blurring using the scattering law of
\citet{2006ApJ...648L.127B} resulted in an inferred intrinsic size of
the source equal to $37^{+16}_{-10}$\,$\mathrm{\mu as}$
\citep{2008Natur.455...78D}.

The inferred image size for \sgra\ was based on fitting sparse
visibility data in the interferometric $u$-$v$ plane with a Gaussian
model (alternate models have also been considered; see
\citealt{2008Natur.455...78D}).
The images from our simulations, however, have significant asymmetry,
either because they are dominated by emission in the funnel or because
of Doppler effects in the disk.
In principle, in order to compare our simulations to the observed
image size, we will need to calculate the predicted scattering
broadened visibilities in the $u$-$v$ plane and compare them directly
to the data (as is done, e.g., in \citealt{2011ApJ...735..110B} and in
\citealt{2009ApJ...703L.142D}).
However, this introduces additional free parameters in the model, such
as the orientation of the black-hole spin vector on the plane of the
sky, and the current coverage of the $u$-$v$ plane is too sparse to
allow us to constrain the model parameters significantly better than
the simple estimate of the size (see, e.g., the large areas within the
confidence contours in \citealt{2011ApJ...735..110B} and in
\citealt{2009ApJ...703L.142D}).
For this reason, we follow a more approximate procedure in comparing
our simulations to the current estimates of the image size at 1.3\,mm.

We take into account the effects of interstellar scattering using the
elliptical scattering kernel of \citet{2006ApJ...648L.127B} that has a
major axis
\begin{align}
  \mathrm{FWHM}_\mathrm{major} = 1.309(\lambda/1\,\mathrm{cm})^2\,\mathrm{mas},
\end{align}
and a minor axis
\begin{align}
  \mathrm{FWHM}_\mathrm{minor} = 0.64(\lambda/1\,\mathrm{cm})^2\,\mathrm{mas},
\end{align}
with a position angle of the major axis at $\mathrm{PA} = 78^\circ$
East of North.
Because we do not know a priori the position angle of the black-hole
spin axis, we consider the entire range of relative position angles
between the major axis of the scattering kernel and the spin axis,
$\DPA$.
For each configuration, we convolve the ray traced image from the
simulation with the scattering kernel and fit the resulting blurred
image with a single Gaussian profile.

\begin{figure*}
  \includegraphics[width=\textwidth]{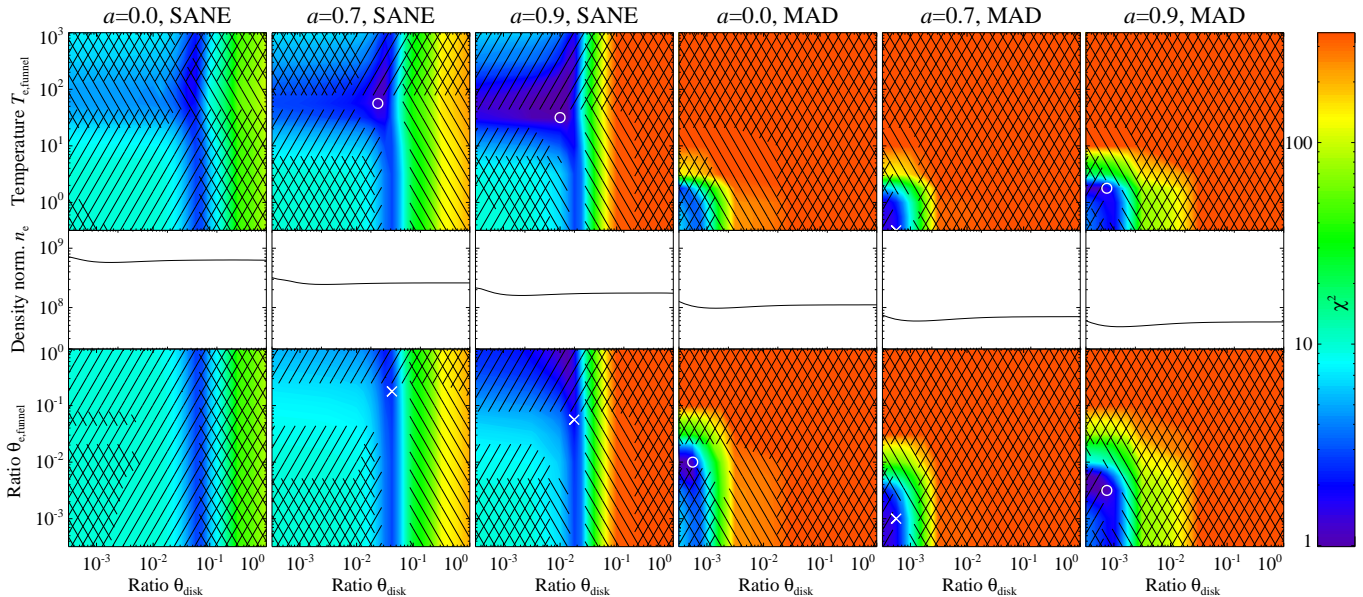}
  \caption{Constraints on model parameters obtained by comparing the
    results of simulations to the observed spectra and image sizes of
    \sgra, as described in the text and shown in Figures~\ref{fig:spec}.
    Each column above corresponds to a different GRMHD simulation,
    while the top and bottom panels correspond to the two different
    descriptions of the plasma thermodynamics in the funnel.
    The solid curves in the middle row depict the electron density
    normalization for which the predicted X-ray fluxes agree with 10\%
    of the observed quiescent flux.
    The colored contour plots show values of $\chi^2$ (see the color
    bar) obtained from fitting the simulated spectra to the
    radio/sub-mm data.
    The forward slanted lines ``///'' show the regions in the
    parameter space that are excluded because the predicted infrared
    flux lies outside of the observed limits.
    Finally, the backward slanted lines
    ``$\backslash\backslash\backslash$'' show the constraints on the
    image size at 1.3\,mm.
    The best-fit set of parameters for each simulation is obtained and
    is shown as a white circle or a white cross.
    The white circles are for fits with $\chi^2 < 1.5$, which we
    consider to be good fits; the white crosses are for $\chi^2 > 1.5$.
    Note that for the \texttt{a0SANE} models and the constant
    temperature \texttt{a0MAD} model, no set of parameters satisfies
    all of the observed constraints.}
  \label{fig:fit}
  \vspace{6pt}
\end{figure*}

In Figure~\ref{fig:dpa}, we show how the relative angle $\DPA$ between
the black-hole spin and the major axis of the scattering kernel
affects the blurred image sizes.
The top row shows 1.3\,mm images for the \texttt{a0SANE} simulation,
using the constant ratio model; the bottom row shows 1.3\,mm images
for the \texttt{a9SANE} simulation using the constant temperature
model.
Note that, in anticipation of the results presented in the following
section, the plasma parameters for these two sets of images correspond
to the best fit values for \sgra, which we are going to obtain in the
following section.

In each row, the first column shows the results of the direct ray
tracing simulations.
The remaining columns show the same image after taking into account
the effects of interstellar scattering, for three values of the
relative position angles.
The dependence of the inferred image size on this parameter is weak.
Moreover, in some cases, because of the asymmetries in the images, the
blurred image size is not a monotonic function of the relative
position angle.
For this reason, for each simulation and for each set of model
parameters, we scan the entire range of relative orientation angles
and consider the range of predicted sizes as an uncertainty in the
model.
We then compare the predicted values to the observed image size of
$43^{+14}_{-8}\mathrm{\mu as}$.

\section{Constraints on Sgr~A*}
\label{sec:fit}

For each of the GRMHD simulations, which corresponds to different
black-hole spins and magnetic field configurations, and for either the
constant ratio or the constant temperature models, we use observations
described in the last section to constrain the five additional model
parameters.
In particular, we use, \emph{(i)} the quiescent flux in the X-rays,
\emph{(ii)} the observed flux and spectral shape in the frequency
range $10^{11}\;\mathrm{Hz}\le \nu \le 10^{12}\;\mathrm{Hz}$,
\emph{(iii)} the range of observed fluxes at $\nu \approx 1.3\times
10^{14}\;\mathrm{Hz}$, and \emph{(iv)} the image size at 1.3\,mm.
This is a large parameters space, which is difficult to explore
computationally and may lead to large degeneracies between model
parameters.

Following \citet{2013A&A...559L...3M}, we fix the threshold
plasma-$\beta$ value that separates the disk from the funnel to
$\Bt=0.2$.
Changing the value of this parameter would simply lead to correlated
changes in the electron-to-ion temperature ratio in the disk, without
affecting significantly the overall results.
We take advantage of the fact that the spectra are insensitive to the
inclination angle at low frequencies (see the third row of
Figure~\ref{fig:var}) and set $i = 60^\circ$ \citep[this value was
  justified by][and will be checked later for consistency]{Psaltis2014}.
Finally, as we discussed in \S\ref{sec:properties}, several aspects of
the simulated spectra and images depend very weakly on some of the
model parameters.
This allows us to follow the procedure described below, which leads us
to use different aspects of the observations to constrain successive
subsets of the remaining model parameters.

\emph{Fixing the electron density scale using the X-ray flux.---\/}In
Figure~\ref{fig:var}, we showed that, among the five model parameters,
the predicted X-ray flux at $\nu \approx 10^{18}$~Hz for each
simulation is only sensitive to the density normalization $\Ne$ when
we fix the inclination.
This property lets us find a correlation between $\Ne$ and $\td$ (with
$\tf = \Tf = 0$) such that the simulated X-ray flux agrees with the
observed quiescent flux.
The result is shown as a set of solid curves in the second row of
Figure~\ref{fig:fit}.
Note that, because the accretion rates, in code units, are comparable
for all simulations \citep[see top panel of Figure~4
  in][]{2013MNRAS.436.3856S}, the density normalizations shown here
indicate that \texttt{a0SANE} has the highest physical accretion rate,
while \texttt{a9MAD} has the lowest.

\emph{Fixing the electron temperature in the disk using the mm-to-cm
  spectrum.---\/}For each value of the electron-to-ion temperature
ratio in the disk, the observed quiescent flux in the X-rays sets the
electron density scale.
Given that we have also fixed the values of two parameters, i.e., $i =
60^\circ$ and $\Bt = 0.2$, the predicted mm-to-cm spectrum of
\sgra\ is now only a function of the remaining two parameters, which
are the electron-to-ion temperature ratio in the disk, $\td$, and
either the constant electron density in the funnel $\Tf$ or the
electron-to-ion temperature ratio in the funnel $\tf$, depending on
which plasma model we are considering.
For each GRMHD simulation and for each plasma model, we show contours
of $\chi^2$ values obtained by comparing the model prediction to the
observed mm-to-cm spectrum of \sgra.
In each panel, dark blue/purple colors represent small $\chi^2$ values
(see the color bar) and hence are better fits.

\emph{Ruling out bright funnels using the infrared flux.---\/}In
Figure~\ref{fig:var}, we showed that the predicted flux at infrared
wavelengths is sensitive to the thermodynamic properties of the
electrons in the funnel, with a weaker dependence on the electron
temperature in the disk.
In particular, when the electron temperature or the electron-to-ion
temperature ratio in the funnel become large (and the funnel becomes
very bright), the predicted infrared flux becomes too large to account
for the range of observed infrared fluxes from \sgra.
The same is true for the electron-to-ion temperature in the disk.
In Figure~\ref{fig:fit}, we use forward slanted lines ``///'' to show
the regions of the parameter space that generate infrared fluxes
outside of the observed range.

\begin{deluxetable}{lccccrc}
  \tablewidth{\columnwidth}
  \tablecaption{The Best-Fit Parameters for Different Simulations}
  \tablehead{Model & $\Ne(\td)$ & $\td$ & $\Tf$ & $\tf$ & $\chi^2$}
  \startdata
  \texttt{a7SANE} & $6.885\times10^7$ & 0.02371 & 56.23  & ---   & 1.178 & $\circ$ \\
  \texttt{a7SANE} & $6.940\times10^7$ & 0.04217 & ---  & 0.17783 & 2.577 & $\times$ \\
  \texttt{a9SANE} & $5.465\times10^7$ & 0.01000 & 31.62  & ---   & 0.674 & $\circ$ \\
  \texttt{a9SANE} & $5.561\times10^7$ & 0.01778 & ---  & 0.05623 & 1.829 & $\times$ \\
  \texttt{a0MAD}  & $5.932\times10^8$ & 0.00056 & ---  & 0.01000 & 1.202 & $\circ$ \\
  \texttt{a7MAD}  & $2.495\times10^8$ & 0.00056 & \ 0.32 & ---   & 1.721 & $\times$ \\
  \texttt{a7MAD}  & $2.495\times10^8$ & 0.00056 & ---  & 0.00100 & 1.791 & $\times$ \\
  \texttt{a9MAD}  & $1.599\times10^8$ & 0.00075 & \ 1.78 & ---   & 1.471 & $\circ$ \\
  \texttt{a9MAD}  & $1.599\times10^8$ & 0.00075 & ---  & 0.00316 & 0.922 & $\circ$
\enddata

  \tablecomments{The above nine fits have $\chi^2$ range from $\sim
    0.5$ to $\sim 2.5$.
    The five fits with $\chi^2 < 1.5$ are equally good matches to the
    observational data of \sgra, which correspond to the white circles
    in Figure~\ref{fig:fit}.
    The four fits with $\chi^2 > 1.5$, correspond to the white crosses
    in Figure~\ref{fig:fit}, match the observation less well.
    The remaining three configurations, namely, \texttt{a0SANE} with
    the two different plasma models and \texttt{a0MAD} with constant
    temperature model, do not contain any set of parameters that can
    simultaneously satisfy all the observational constraints.}
  \label{tab:fit}
\end{deluxetable}

\emph{Rejecting models using the 1.3\,mm image sizes.---\/}The final
constraint on the model parameters arises from the comparison of the
predicted to the observed 1.3\,mm image sizes.
We use backward slanted lines ``$\backslash\backslash\backslash$'' to
indicate the parameter space that is excluded by this constraint.
With this final constraint folded in, we find the best-fit set of
parameters for each simulation within the allowed regions of the
parameter space and mark them with white circles or crosses in
Figure~\ref{fig:fit} and list them in Table~\ref{tab:fit}.
The white circles in Figure~\ref{fig:fit} are for fits with $\chi^2 <
1.5$, which we consider to be good fits; the white crosses are for
$\chi^2 > 1.5$.
Note that for the \texttt{a0SANE} models and the constant temperature
\texttt{a0MAD} temperature model, no set of parameters satisfies all
of the observed constraints.

\begin{figure}
  \includegraphics[width=\columnwidth,trim=12 6 12 12]{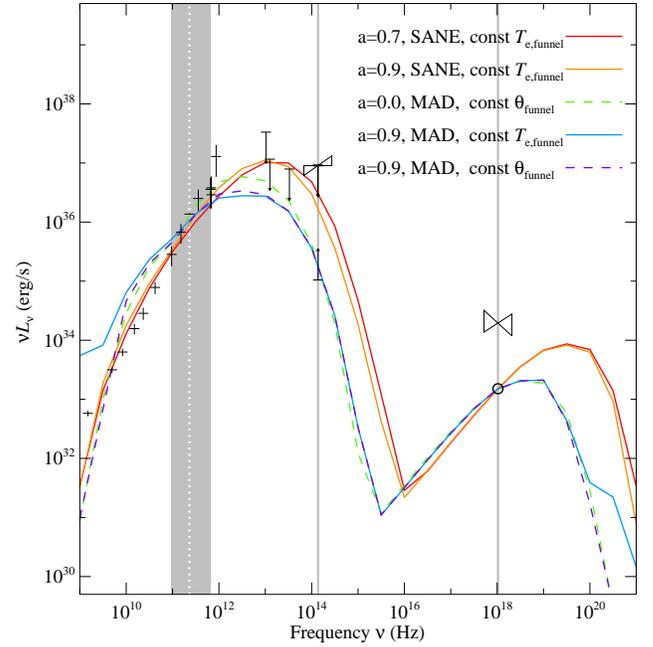}
  \caption{The five best-fit modeled spectra and the broadband
    spectral data that we used.
    The gray band between $\nu \approx 10^{11}\,\mathrm{Hz}$ and
    $\approx 10^{12}\,\mathrm{Hz}$ marks the frequency range over
    which we perform the least-squares fits.
    The gray line at $\nu \approx 10^{14}\,\mathrm{Hz}$ marks the
    infrared frequency at which we used the range of fluxes observed
    at different times to impose an upper and lower bound on the models.
    The gray line at $\nu \approx 10^{18}\,\mathrm{Hz}$ marks the
    X-ray frequency, at where we used 10\% of the observed quiescent
    flux (the open circle below the bow-tie) to fix the density
    normalization in the flow.
    The white dotted line inside the gray band marks $\lambda =
    1.3\,\mathrm{mm}$, where we used the EHT measurement of the image
    size.
    The model parameters are marked as white circles in
    Figure~\ref{fig:fit} and listed in Table~\ref{tab:fit}.}
  \label{fig:spec}
  \vspace{6pt}
\end{figure}

\begin{figure*}
  \includegraphics[width=\textwidth,trim=0 0 0 0]{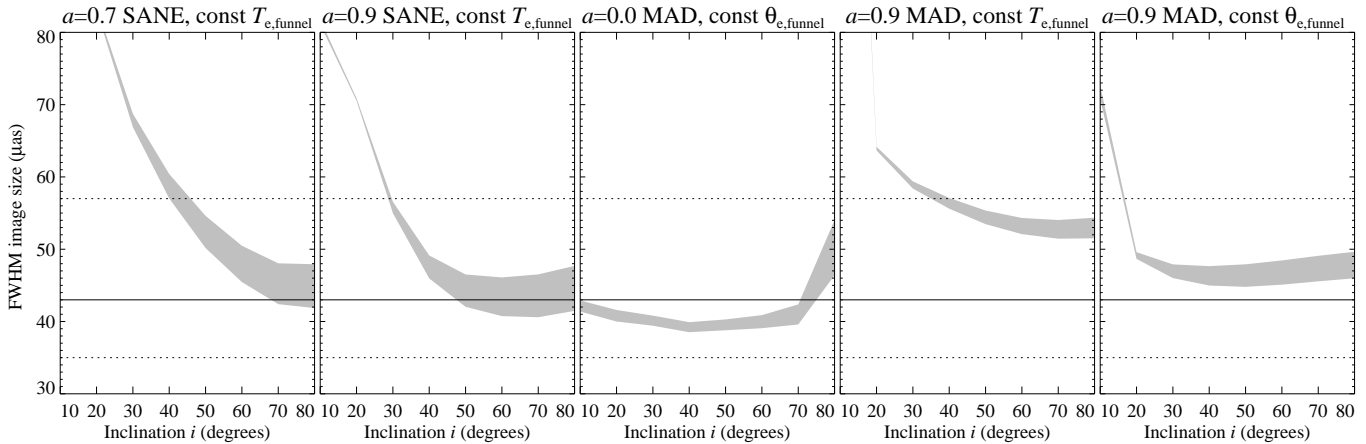}
  \caption{The dependence of image sizes on the inclination angle $i$
    for the five best-fit models listed in Table~\ref{tab:fit}.
    For each inclination, we sample a range of values for the relative
    position angle between the major axis of the scattering kernel and
    the black-hole spin, $\DPA = 0^\circ$, $10^\circ$, $20^\circ$,
    \dots, $170^\circ$, in order to obtain upper and lower bounds on
    the predicted sizes and show their range as a gray band.
    The solid lines mark the image size (including interstellar
    scattering) obtained at 1.3\,mm from EHT observations and the
    dotted lines mark uncertainties at the $3\sigma$ level.
    While the image sizes decrease for high inclination for most
    simulations, it increases for high inclination for the
    non-spinning MAD simulation \texttt{a0MAD}.
    The image sizes for \texttt{a9MAD} jump to large values for $i =
    10^\circ$ because, at small inclinations, the observer looks down
    along the strong funnel.}
  \label{fig:size}
  \vspace{6pt}
\end{figure*}

In Figure~\ref{fig:spec}, we plot the best-fit model spectra together
with the broadband spectral data that we use for the fitting.
The gray band between $\nu \approx 10^{11}\,\mathrm{Hz}$ and $\approx
10^{12}\,\mathrm{Hz}$ marks the frequency range over which we perform
the least-squares fits.
The gray line at $\nu \approx 10^{14}\,\mathrm{Hz}$ marks the infrared
frequency at which we use the range of fluxes observed at different
times to impose an upper and lower bound on the models.
The gray line at $\nu \approx 10^{18}\,\mathrm{Hz}$ marks the X-ray
frequency where we used 10\% of the observed quiescent flux (the open
circle below the bow-tie) to fix the density normalization in the
flow.
The white dotted line inside the gray band marks $\lambda =
1.3\,\mathrm{mm}$, where we use the EHT measurement of the image size.
The five curves correspond to the best-fit models with parameters
marked as white circles in Figure~\ref{fig:fit} and listed in
Table~\ref{tab:fit}.
For these fits, the two SANE simulations have spectral peaks near $\nu
\approx 10^{13}\;\mathrm{Hz}$ with $\nu L_\nu \approx
10^{37}\;\mathrm{erg/s}$, and are indistinguishable from each other.
The remaining three MAD simulations have spectral peaks near $\nu
\approx 3\times10^{12}\;\mathrm{Hz}$ with half an order of magnitude
lower fluxes at these frequencies.

As a final consistency check, we plot in Figure~\ref{fig:size} the
range of image sizes for the best-fit models as functions of the
observer's inclination.
As discussed in \citet{Psaltis2014}, in models that are dominated by
disk emission, the inclination affects primarily the size of the
1.3\,mm image, because of the Doppler effect.
This is clearly seen in the first two panels that correspond to the
SANE simulations.
However, the dependence of the 1.3\,mm image size on inclination for
the \texttt{a0MAD} simulation is very weak, with a size that actually
increases with incliation.
In all cases, the value of $i=60^\circ$ that we have adopted for the
inclination of the observer is consistent with observations.

\section{Conclusions}
\label{sec:discussions}

\begin{figure}[h]
  \includegraphics[width=\columnwidth,trim=0 0 0 0]{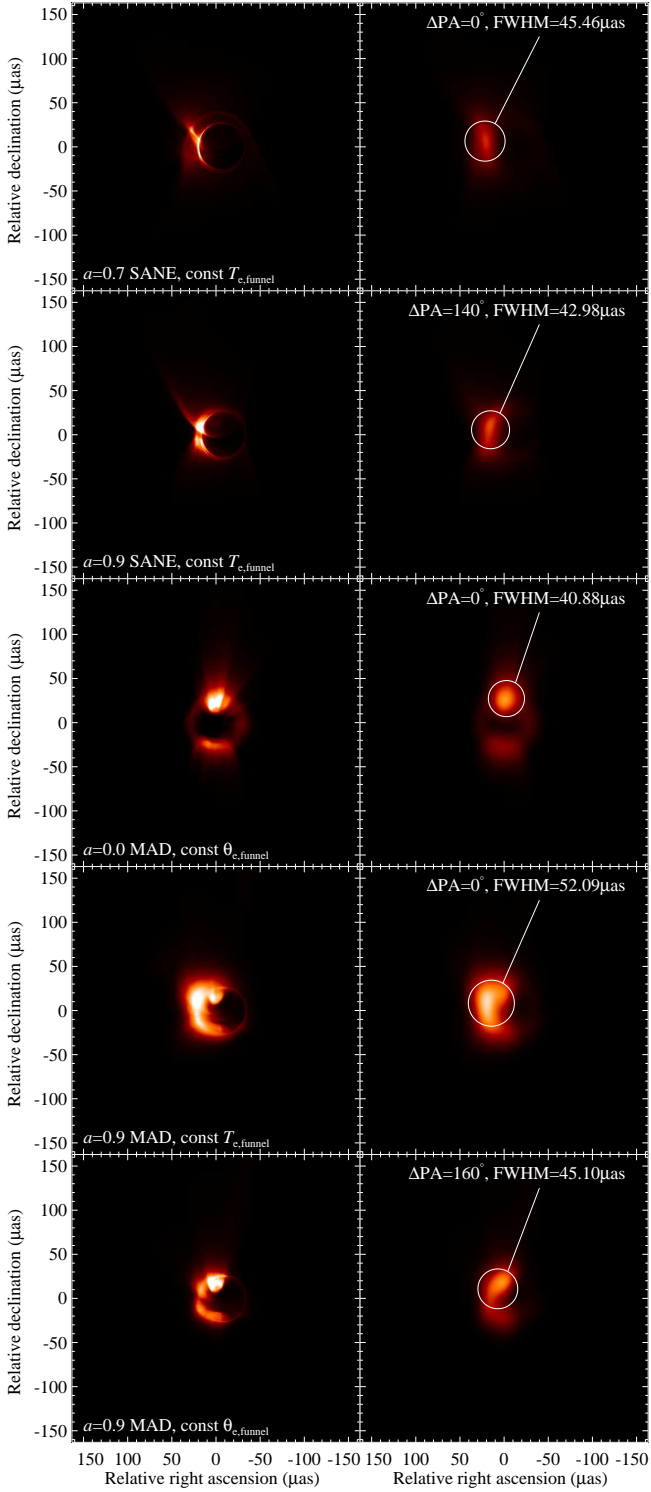}
  \caption{Predicted EHT images at $\lambda = 1.3\,\mathrm{mm}$, for
    the five best fit models shown with white circles in
    Figure~\ref{fig:fit}.  The left column shows the results of the
    direct ray-tracing simulations, while the right column shows the
    scatter-broadened images.  In the disk-dominated SANE simulations,
    the images have the characteristic crescent shape of a Doppler
    boosted accretion flow and a clear imprint of the black-hole
    shadow.  In the jet-dominated MAD simulations, the image is formed
    primarily by the emission in the jet footprints but the shadow is
    still visible.}
  \label{fig:img}
\end{figure}

In this paper, we analyzed the predicted spectra and images of various
GRMHD simulations and plasma models for the accretion flow around
\sgra, and narrowed the model parameter range by imposing the
requirement that our predictions be consistent with the observed
spectra and the 1.3\,mm image size of \sgra.  As seen in
Figure~\ref{fig:spec}, the best fit model spectra agree with
observational data very well.  The models are most variable at
infrared wavelengths, which is also consistent with observations.  It
is important to emphasize here that, if the spectral properties of
\sgra\ were considered without any image size constraints, the range
of allowed model parameter would be significantly wider.  This
demonstrates the power of using images with horizon-scale resolution
to distinguish between models that would otherwise make seemingly
similar predictions.

Future imaging observations from the completed EHT is expected to more
easily distinguish disk-dominated from all funnel-dominated models.
In Figure~\ref{fig:img}, we show the predicted 1.3\,mm images for
\sgra\ for the GRMHD simulations and plasma model parameters that are
consistent with all current spectra and imaging observations.  With
the completed EHT, it will be straightforward to distinguish between
the first two pairs of images from the remaining three pairs of images
shown, i.e., the disk-dominated from the funnel models, respectively.
However, distinguishing among the various disk-dominated models and
measuring the parameters of the black-hole and of the plasma, will
require additional information from the EHT, including the
polarization and scale dependent variability.  We will explore these
aspects of our models in future work.

\acknowledgments

This research is conducted at the University of Arizona and is
supported by NSF grant AST~1312034.
The Arizona team and R.~N. acknowledge the support of NASA/NSF TCAN
award NNX14AB48G.
All ray tracing calculations were performed in the \texttt{El~Gato}
GPU cluster that is funded by NSF award 1228509.
We thank Alexander Tchekhovskoy for providing the original derivations
of the coordinate transformation used in the HARM simulations.
We also thank Eliot Quataert and Jason Dexter for numerous
discussions.
F.~\"O. thanks the Miller Institute for Basic Research in Science at
the University of California Berkeley for their support and
hospitality.

\bibliography{my,ms}

\end{document}